\documentclass[12pt]{article}

\catcode`\@=11
\global\arraycolsep=2pt
\usepackage{amsbsy,amssymb,latexsym,amsfonts,amsmath}
\usepackage{graphicx,color}

\begin{document}
\begin{flushright}
\parbox{4.2cm}
{CALT 68-2935}
\end{flushright}

\vspace*{0.7cm}

\begin{center}
{\Large Holographic interpretation of renormalization group approach to singular perturbations in non-linear differential equations}
\vspace*{1.5cm}\\
{Yu Nakayama}
\end{center}
\vspace*{1.0cm}
\begin{center}
{\it California Institute of Technology,  \\ 
452-48, Pasadena, California 91125, USA}
\vspace{3.8cm}
\end{center}

\begin{abstract}
We give a holographic explanation how the renormalization group approach to singular perturbations in non-linear differential equations proposed by Chen, Goldenfeld and Oono is indeed equivalent to a renormalization group method in quantum field theories proposed by Gell-Mann and Low via AdS/CFT correspondence.
\end{abstract}

\thispagestyle{empty} 

\setcounter{page}{0}

\newpage

The philosophy of the renormalization group (e.g. \cite{Wilson:1973jj} for a review) shows far richer applications in mathematical physics beyond the original scope of quantum field theories and many body systems. In particular, it has a dramatic application in asymptotic analysis of certain non-linear differential equations as first advocated by Chen, Goldenfeld and Oono \cite{Chen:1994zza}\cite{Chen:1995ena}. While the philosophy of the renormalization group in asymptotic analysis has demonstrated a beautiful universal structure, a direct connection to the  renormalization group method developed in quantum field theories and many body systems was lacking.

In this paper, we propose a field theory interpretation of Chen-Goldenfeld-Oono renormalization group in certain non-linear differential equations from AdS/CFT correspondence (e.g. \cite{Aharony:1999ti} for a review). It will turn out that it is precisely the holographic realization of the Gell-Mann Low renormalization group equation \cite{GL} of the dual quantum field theory. Our physical realization is motivated by the radion stabilization problem studied in \cite{Chacko:2013dra}\cite{Bellazzini:2013fga}, and it gives a more transparent viewpoint of their singular perturbation theory with the boundary layer analysis.

Let us consider the $\mathrm{AdS}_5$ space-time whose metric in Poincar\'e patch is
\begin{align}
ds^2 = g_{MN} dx^M dx^N = dr^2  +e^{-2Ar} \eta_{\mu\nu} dx^{\mu}dx^{\nu} \ ,
\end{align}
where $\eta_{\mu\nu} = \mathrm{diag}(-1,1,1,1)$.
In applications of AdS/CFT correspondence, we are often interested in the boundary value problems for various source fields living in the bulk. We put the boundary at $r= r_{\mathrm{UV}}$ and $r= r_{\mathrm{IR}}$ so that the bulk is defined by the range $r_{\mathrm{UV}} < r < r_{\mathrm{IR}}$. It is customary to take $e^{-2A(r_{\mathrm{UV}} - r_{\mathrm{IR}})} \gg 1$ to achieve a large hierarchy. It is important to recall that the fifth coordinate $r$ is related to the renormalization scale $\mu$ of the dual quantum field theory  via $\log \mu = Ar$.

For concreteness, we focus on a scalar field in the $\mathrm{AdS}_5$ space-time whose action is
\begin{align}
 S = \int d^{5} x \sqrt{|g|} \left(\frac{1}{2}\partial_M \Phi \partial^M \Phi + V(\Phi) \right) \ . \label{saction}
\end{align}
The potential $V(\Phi)$ can be arbitrary
\begin{align}
V(\Phi) = \frac{1}{2}m^2 \Phi^2 + \frac{1}{3!}\eta \Phi^3 + \frac{1}{4!}\zeta \Phi^4 + \cdots \ ,
\end{align}
but in order to connect to relevant but a nearly marginal deformation of the ultraviolet dual conformal field theory, we take $m^2 <0$ with $\frac{|m^2|}{A^2} \ll 1$.
Furthermore, we will consider the probe limit and neglect the back-reaction to the metric in the following to extract the essence of our discussions. 

The bulk equation of motion for the scalar field from the action \eqref{saction} is 
\begin{align}
\partial_r^2 \Phi(r) - 4A \partial_r \Phi(r) - V'(\Phi(r)) = 0 \ , \label{eom}
\end{align}
and one of the typical problems in AdS/CFT correspondence is to solve the equation \eqref{eom} with a specified boundary condition $\Phi(r_{\mathrm{UV}}) = \Phi_{\mathrm{UV}}$ and $\Phi(r_{\mathrm{IR}}) = \Phi_{\mathrm{IR}}$. Such analysis is important in studying correlation functions, linear response theory, radion stabilization and so on in the context of AdS/CFT correspondence.
Other boundary conditions such as $\partial_{r} \Phi(r_{\mathrm{IR}}) = \Phi'_{\mathrm{IR}}$ are equally possible, but details will not affect the following argument at all. 

Let us illustrate how the renormalization group approach to singular perturbations in non-linear differential equations proposed by Chen, Goldenfeld and Oono is applicable here in the simplest case with $V(\Phi) = \frac{1}{2}m^2\Phi^2$. Of course,  the equations of motion \eqref{eom} from the quadratic action is trivially solvable with
\begin{align}
\Phi^{\mathrm{exact}}(r) = C e^{\alpha_- r} + \tilde{C} e^{\alpha_+ r} \ ,
\end{align}
where $\alpha_{\pm} = 2A \pm \sqrt{4A^2 + m^2}$, and $C$, $\tilde{C}$ are determined by the boundary condition,
but we follow the Venus physicist approach to draw a lesson. Let us treat $\frac{m^2}{A^2}$ as a perturbation with respect to the zeroth order solution $\Phi_0 = C_0 + \tilde{C}_0 e^{4rA}$.  
The naive first order perturbation would lead to
\begin{align}
\Phi_1^{\mathrm{naive}}(r) = C_0 \left(1-\frac{m^2}{4A} r\right) + \tilde{C}_0 e^{4rA}\left(1 + \frac{m^2}{4A}r\right) + \mathcal{O}\left(\frac{m^4}{A^4}\right) \ . 
\end{align}
The naive perturbation theory breaks down for large (negative) $r$ because the corrections are secular terms. Also it does not respect the natural shift symmetry of $r$.

The idea to resolve the problem with the singular perturbation with secular terms is to sum up the ``leading log" or to use the renormalization group method. We introduce the floating cut-off $r_0$ and set $r = r + r_0 -r_0$ and renormalize the initial condition $C_0$ and $\tilde{C}_0$ by absorbing the ``logarithmic divergence" from $r_0 = \frac{1}{A} \log \mu_0$. 
The naive perturbation series can be recast into the renormalized series:
\begin{align}
\Phi_1^{\mathrm{ren}}(r) = C_0(r_0) \left(1-\frac{m^2}{4A}(r -r_0) \right) + \tilde{C}_0 (r_0) e^{4rA} \left(1+ \frac{m^2}{4A}(r-r_0) \right) + \mathcal{O}\left(\frac{m^4}{A^4}\right) \ .
\end{align}
Since the physical observable $\Phi(r)$ cannot depend on the floating cut-off $r_0$ which is arbitrary, we demand the renormalization group equation \cite{Chen:1994zza} $\frac{\partial}{\partial r_0} \Phi_1^{\mathrm{ren}}(r) = 0$, or
\begin{align}
\frac{d}{d r_0} C_0(r_0) = -\frac{m^2}{4A} C_0(r_0) \ , \ \ \frac{d}{d r_0} \tilde{C}_0(r_0)  = \frac{m^2}{4A} \tilde{C}_0(r_0) \ . \label{CGOren}
\end{align}
By solving the renormalization group equation and setting $r_0 = r$ (for the best approximation compatible with the first order perturbation), we obtain the first order approximate solution with the renormalization group improvement:
\begin{align}
\Phi^{\mathrm{ren}}_1(r) = C^{\mathrm{ren}}_0 e^{-\frac{m^2}{4A} r } + \tilde{C}^{\mathrm{ren}}_0 e^{4Ar + \frac{m^2}{4A} r} + \mathcal{O}\left(\frac{m^4}{A^4}\right) \ .
\end{align}
In this case, the renormalization condition $r_0 =r$ also removes the secular terms in the naive perturbative solution. At this point, the renormalized constant $C_0^{\mathrm{ren}}$ and $\tilde{C}_0^{\mathrm{ren}}$ can be determined by the boundary condition we impose. Morally speaking, the ultraviolet data specifies $C_0^{\mathrm{ren}}$ while the infrared data specifies $\tilde{C}_0^{\mathrm{ren}}$ as long as we have the large hierarchy $e^{-2A(r_{\mathrm{UV}} - r_{\mathrm{IR}})} \gg 1$.

One should note that this approach reproduces the singular perturbation theory with the boundary layer analysis used in \cite{Chacko:2013dra}. Moreover it gives systematic corrections in higher orders. The most important point to realize, however, is that the renormalization group equation \eqref{CGOren} is precisely the holographic renormalization group equation obtained  in the AdS/CFT literature. 

We recall that the boundary value of the scalar field $\Phi(r=r_{\mathrm{UV}})$ is identified with the (ultraviolet) coupling constant $g_{\mathrm{UV}}$ of the dual quantum field theory. Within the holographic scheme \cite{de Boer:1999xf}\cite{Anselmi:2000fu}, it is natural to extract the field theory beta function from the scaler field in the bulk from the identification
\begin{align}
\beta(g(\mu)) \equiv \frac{1}{A} \frac{d\Phi}{dr} 
\end{align}
with $\log \mu = Ar$ in mind. Near the ultraviolet boundary, the Chen-Goldenfeld-Oono renormalization group with the holographic interpretation gives $\beta(g) \sim -\frac{m^2}{4A^2} g$, and it reproduces the conformal perturbation theory result of the dual quantum field theory with the Gell-Mann Low renormalization group equation:
\begin{align}
\frac{dg(\mu)}{d\log\mu} = \beta(g(\mu)) = (4-\Delta) g(\mu) + \mathcal{O}(g^2) \ .
\end{align}
Here $\Delta$ is the scaling dimension of the perturbing operator and it is given by $4-\Delta \sim -\frac{m^2}{4A^2}$ via the standard AdS/CFT dictionary $\Delta_{\pm} = 2 \pm \sqrt{4+\frac{m^2}{A^2}}$.  Therefore, the renormalization group equation \eqref{CGOren} is equivalent to the renormalization group invariance of the observed coupling constant of the dual quantum field theory.

The above argument can be easily generalized to the non-linear situations with the generic potential $V(\Phi)$. The dangerous secular term is renormalized by adjusting the initial condition $C_0$, and the renormalization group equation ala Chen-Goldenfeld-Oono 
\begin{align}
\frac{d}{d r_0} C_0(r_0) = - \frac{V'(C_0(r_0))}{4A}  \ 
\end{align}
is identical to the holographic renormalization group equation in the leading order approximation (more precisely when the superpotential and potential can be identified within the probe approximation). With the AdS/CFT correspondence, we can interpret it as the Gell-Mann Low renormalization group equation of the dual quantum field theory.

As pointed out in \cite{Chen:1995ena}, we may also derive the renormalization group equation from the Wilsonian viewpoint. The naive perturbation is more trustful for infinitesimal change $\delta r$ than the one-time integration over the large scale $A (r_{\mathrm{IR}}-r_{\mathrm{UV}}) \gg 1$.
Thus, starting with the constant unperturbed solution $\Phi_0(r) = C_0$, we obtain the infinitesimal integration 
\begin{align}
\Phi_1 (r+ \delta r) = \Phi_1(r)\left(1 -\frac{\delta r}{4A} \frac{V'(\Phi_1(r))}{\Phi_1(r)} \right) \ 
\end{align}
so that we can set up the Wilsonian-type renormalization group equation
\begin{align}
\frac{d\Phi_1(r)}{dr} = -\frac{V'(\Phi_1(r))}{4A}
\end{align}
 by successively renormalizing the initial condition
as a better starting point for the perturbative computation without dangerous large log corrections. This is precisely equivalent to the outer region solution \cite{Chacko:2013dra} in the boundary layer analysis.
Translating integration over the large scale at once into the step-by-step differentiation is the key philosophy of the Wilsonian renormalization group. The same philosophy has been pursued in holography to obtain the radial flow of the bulk fields \cite{Akhmedov:2002gq}\cite{Heemskerk:2010hk}\cite{Lee:2012xba}\cite{Balasubramanian:2012hb}\cite{Lee:2013xba}. Our observation provides a novel viewpoint from the asymptotic analysis of the bulk differential equations.
 We could address the similar question in the second unperturbed solution $\Phi_0(r) = \tilde{C}_0 e^{4Ar}$, and we observe that it is related to the renormalization of the vacuum expectation value of the dual operator.

As a concrete non-linear problem, let us revisit the analytically manageable example studied in \cite{Chacko:2013dra}. We consider the $\mathrm{AdS}_5$ boundary value problem with the cubic scalar interaction $V(\Phi) = \frac{1}{3!}\eta \Phi^3$. The naive perturbative computation gives
\begin{align}
\Phi_1^{\mathrm{naive}}(r) = C_0 + \tilde{C}_0 e^{4Ar} - \frac{C_0^2}{8A}\eta r + \frac{C_0 \tilde{C}_0}{4A}\eta e^{4Ar}  r + \frac{\tilde{C}_0^2}{64A^2} \eta e^{8Ar} - \frac{C_0 \tilde{C}_0}{16A^2} \eta e^{4Ar} + \mathcal{O}\left(\frac{\eta^2}{A^4}\right) \ . 
\end{align}
The dangerous secular terms can be removed via renormalizing $C_0$ and $\tilde{C}_0$ by introducing the renormalized initial condition for $C_0(r_0)$ and $\tilde{C}_0(r_0)$ with $r = r + r_0 - r_0$ as before:
\begin{align}
\Phi_1^{\mathrm{ren}}(r) =& C_0(r_0) + \tilde{C}_0(r_0) e^{4Ar} - \frac{C_0(r_0)^2}{8A}\eta (r-r_0) + \frac{C_0(r_0) \tilde{C}_0(r_0) }{4A}\eta e^{4Ar}  (r-r_0) \cr
 &+ \frac{\tilde{C}_0(r_0)^2}{64A^2} \eta e^{8Ar} - \frac{C_0(r_0) \tilde{C}_0(r_0)}{16A^2} \eta e^{4Ar} + \mathcal{O}\left(\frac{\eta^2}{A^4}\right) \ . 
\end{align}

The renormalization group equation  $\frac{\partial}{\partial r_0} \Phi_1^{\mathrm{ren}} (r) = 0$ to eliminate the floating cut-off dependence now yields
\begin{align}
\frac{dC_0(r_0)}{dr_0} = -\frac{C_0(r_0)^2}{8A}\eta \ , \ \ \frac{d\tilde{C}_0(r_0)}{dr} = \frac{C_0(r_0) \tilde{C}_0(r_0)}{4A} \eta  \ . 
\end{align}
The first equation is nothing but the one-loop Gell-Mann Low equation for the marginal deformation in the conformal perturbation theory:
\begin{align}
\frac{dg(\mu)}{d\log\mu} = \beta_0 g(\mu)^2 + \mathcal{O}(g^3) \ , 
\end{align}
 and it has the solution 
\begin{align}
C_0(r_0) = \frac{C_0^{\mathrm{ren}}}{1+\frac{\eta}{8A} C_0^{\mathrm{ren}} r_0} \ ,  \ \ \tilde{C}_0(r_0) = \tilde{C}_0^{\mathrm{ren}} \left(1+ \frac{\eta }{8A} C_0^{\mathrm{ren}} r_0 \right)^2 \ .
\end{align}
Substituting it back into the renormalized perturbation series (by setting $r_0= r$ for the best approximation) reproduces the result in \cite{Chacko:2013dra} from the boundary layer analysis with further corrections:
\begin{align}
\Phi^{\mathrm{ren}}_1(r) = \frac{C_0^{\mathrm{ren}}}{1+\frac{\eta}{8A} C_0^{\mathrm{ren}} r} + \tilde{C}_0^{\mathrm{ren}} \left(1 + \frac{\eta }{8A}  C_0^{\mathrm{ren}} r\right)^2 e^{4Ar} + \cdots \ .
\end{align}
We could systematically evaluate the higher order corrections if desired.
The renormalized constant $C_0^{\mathrm{ren}}$ and $\tilde{C}_0^{\mathrm{ren}}$ can be determined by the boundary condition we impose.

Before conclusion, we note that the renormalization group improvement by itself does not resolve the ``Landau pole problem" in this example. 
In the actual AdS/CFT correspondence, it is usually the back-reaction to the metric that will affect the gravitational dynamics, and it will eventually cause a flow to another conformal fixed point or confinement. It is certainly outside the scope of our perturbation theory within the probe limit discussed here. However, the philosophy of the renormalization group should remain beyond the perturbation theory.

In this paper, we have demonstrated that the AdS/CFT correspondence gives a natural framework to interpret the renormalization group approach to singular perturbations in non-linear differential equations proposed in \cite{Chen:1994zza}\cite{Chen:1995ena}. Their renormalization group equation is nothing but the Gell-Mann Low renormalization group equation of the dual quantum field theory. It is extremely interesting to extend our observation to the other differential equations studied in \cite{Chen:1994zza}\cite{Chen:1995ena} and see if they also have the quantum field theory interpretation via holography.

\section*{Acknowledgements}
We would like to thank Z.~Chacko for his beautiful lecture and related discussions which stimulate this study.
This work is supported by Sherman Fairchild Senior Research Fellowship at California Institute of Technology  and DOE grant DE-FG02-92ER40701

\end{document}